\documentclass[twocolumn,pre,showpacs,superscriptaddress]{revtex4}
\usepackage{bm}
\usepackage{tikz}
\usepackage{graphicx}%
\usepackage{color} 
\usepackage{transparent}
\usepackage{psfrag}
\usepackage{dcolumn}
\usepackage{amsmath}
\usepackage{amssymb}
\usepackage{latexsym}
\usepackage{hyperref} 
\usepackage{booktabs}
\usepackage{latexsym}
\begin{document}

\newcommand{\tikzcircle}[2][red,fill=red]{\tikz[baseline=-0.5ex]\draw[#1,radius=#2] (0,0) circle ;}%
\def\bea{\begin{eqnarray}}
\def\eea{\end{eqnarray}}
\def\beq{\begin{equation}}
\def\eeq{\end{equation}}
\def\f{\frac}
\def\k{\kappa}
\def\e{\epsilon}
\def\ve{\varepsilon}
\def\be{\beta}
\def\D{\Delta}
\def\h{\theta}
\def\t{\tau}
\def\a{\alpha}

\def\cDa{{\cal D}[X]}
\def\cD{{\cal D}[x]}
\def\cL{{\cal L}}
\def\cLo{{\cal L}_0}
\def\cLa{{\cal L}_1}

\def\Re{{\rm Re}}
\def\sj{\sum_{j=1}^2}
\def\rk{\rho^{ (k) }}
\def\rek{\rho^{ (1) }}
\def\cek{C^{ (1) }}
\def\rz{\rho^{ (0) }}
\def\rt{\rho^{ (2) }}
\def\rtb{\bar \rho^{ (2) }}
\def\trk{\tilde\rho^{ (k) }}
\def\trek{\tilde\rho^{ (1) }}
\def\trz{\tilde\rho^{ (0) }}
\def\trt{\tilde\rho^{ (2) }}
\def\r{\rho}
\def\tD{\tilde {D}}

\def\rpl{r_\parallel}
\def\rp{{\bf r}_\perp}

\def\s{\sigma}
\def\kb{k_B}
\def\bF{\bar{\cal F}}
\def\F{{\cal F}}
\def\la{\langle}
\def\ra{\rangle}
\def\nn{\nonumber}
\def\up{\uparrow}
\def\dn{\downarrow}
\def\S{\Sigma}
\def\dg{\dagger}
\def\d{\delta}
\def\p{\partial}
\def\l{\lambda}
\def\L{\Lambda}
\def\G{\Gamma}
\def\o{\Omega}
\def\w{\omega}
\def\g{\gamma}

\def\bv{ {\bf b}}
\def\uv{ {\hat{\bm{u}}}}
\def\rv{ {\bf r}}
\def\vv{ {\bf v}}

\def\jv{ {\bf j}}
\def\jr{ {\bf j}_r}
\def\jd{ {\bf j}_d}
\def\jdd{ { j}_d}
\def\noi{\noindent}
\def\a{\alpha}
\def\d{\delta}
\def\p{\partial} 

\def\la{\langle}
\def\ra{\rangle}
\def\e{\epsilon}
\def\n{\eta}
\def\g{\gamma}
\def\break#1{\pagebreak \vspace*{#1}}
\def\hf{\frac{1}{2}}
\def\na{{\eta}_{\rm ac}}
\def\dac{D_v}
\def\n{{\eta}}
\def\gv{\gamma_v}

\def\dact{\tilde{D}_v}

\title{Self-propulsion with speed and orientation fluctuation: exact computation of moments and dynamical bistabilities in displacement}

\author{Amir Shee}%
\email{amir@iopb.res.in}
\affiliation{Institute of Physics, Sachivalaya Marg, Bhubaneswar 751005, India}
\affiliation{Homi Bhaba National Institute, Anushaktinagar, Mumbai 400094, India}
\author{Debasish Chaudhuri}%
\email{debc@iopb.res.in}
\affiliation{Institute of Physics, Sachivalaya Marg, Bhubaneswar 751005, India}
\affiliation{Homi Bhaba National Institute, Anushaktinagar, Mumbai 400094, India}

\date{\today}
             
\begin{abstract}    
{We consider the influence of active speed fluctuations on the dynamics of a $d$-dimensional active Brownian particle performing a persistent stochastic motion. We use the Laplace transform of the Fokker-Planck equation to obtain exact expressions for time-dependent dynamical moments. Our results agree with direct numerical simulations and show several dynamical crossovers determined by the activity, persistence, and speed fluctuation.
The persistence in the motion leads to anisotropy, with the parallel component of displacement fluctuation showing sub-diffusive behavior and non-monotonic variation. 
The kurtosis remains positive at short times determined by the speed fluctuation,    
crossing over to a negative minimum at intermediate times governed by the persistence  before vanishing asymptotically. The probability distribution of particle displacement obtained from numerical simulations in two-dimension shows two crossovers between contracted and expanded trajectories via two bimodal distributions at intervening times. 
While the speed fluctuation dominates the first crossover, the second crossover is controlled by persistence like in the worm-like chain model of semiflexible polymers. 
}         
\end{abstract}

\maketitle

\section{Introduction}

Active particles self-propel, consuming and dissipating internal or ambient energy~\cite{Vicsek2012, Romanczuk2012}. They are driven out of equilibrium at the level of individual elements, breaking the detailed balance condition and the equilibrium fluctuation-dissipation relation. Natural examples of active matter span various length scales, including  motor proteins, motile cells, bacteria, developing tissues,  bird flocks, fish school, and animal herds~\cite{Astumian2002, Reimann2002, Berg1972, Niwa1994, Ginelli2015, Marchetti2013}. Inspired by such biological examples, several artificial active elements were fabricated, e.g., vibrated rods, colloidal swimmers, and asymmetric disks~\cite{Marchetti2013, Bechinger2016}. Active colloids can use diffusiophoresis, electrophoresis, and the Marangoni effect to generate self-propulsion~\cite{Bechinger2016}. 

Due to their non-equilibrium nature, active particles show many remarkable properties strikingly different from their equilibrium counterparts. Experimental and theoretical studies gave significant insight into collective motion, flocking, and motility-induced phase separation~\cite{Marchetti2013, Bechinger2016, Bar2020}. 
Even a single active particle can show rich and counter-intuitive physical properties. In this context, studies of simple models have been crucial. They displayed several ballistic-diffusive crossovers, non-Boltzmann steady-state, localization away from potential minima, and associated re-entrant transition for steady-state properties of trapped particles~\cite{Das2018, Kurzthaler2018b, Malakar2018, Basu2018a, Peruani2007, Basu2019, Majumdar2020, Wagner2017, Elgeti2015, Dhar2019, Shee2020, Santra2020a, Pototsky2012, Malakar2020, Chaudhuri2021}.

Fluctuations are inherent to self-propulsion, with its source and nature varying from system to system. For example, ATP hydrolysis in motor proteins or the chemical reaction in the diffusiophoresis of platinum-gold nano-particles immersed in hydrogen peroxide is inherently stochastic. 
The inbuilt structural asymmetry in Janus colloids determines their instantaneous heading direction of motion, which undergoes orientational fluctuations~\cite{Bechinger2016}. They are often modeled as active Brownian particles (ABP) performing a continuous-time persistent random walk, assuming a constant active speed~\cite{Howse2007,Palacci2010, Redner2013, Fily2014, Sevilla2014, Basu2018}. 
However, the mechanism of active speed generation itself is stochastic. For example, the speed distribution in the run and tumble motion of Myxobacteria is broad ~\cite{Wu2009, Theves2013}, and in the pathogenic  E. coli, it displays a bimodality with peaks corresponding to run and stop~\cite{PerezIpina2019, Otte2021}. This necessitates a description of ABP motion in the presence of speed fluctuations.

In theoretical models, self-propulsion mechanisms can be incorporated in various ways. The energy-depot model is described using a stochastic energy gain and dissipation with a part of dissipated energy leading to self-propulsion~\cite{Schweitzer1998}. Similarly, coupling internal chemical processes with physical movement leads to a Langevin description of self-propulsion in apolar and polar particles~\cite{Dadhichi2018, Ramaswamy2017}.  Consideration of a lattice-based model with an internal chemical process generating self-propulsion led to a continuum description similar to the ABP model, apart from the appearance of additional Gaussian noise in active speed~\cite{Pietzonka2018, Pietzonka2019, Pietzonka2016}.

In this paper, we consider the impact of such active speed fluctuations in the dynamics of ABPs. We utilize a Laplace transform approach initially developed to understand the properties of the worm-like chain (WLC) model of semiflexible polymers~\cite{Hermans1952} to calculate the exact time dependence of all moments of the ABP in arbitrary dimensions from the Fokker-Planck equation. 
We calculate the time dependence of  mean-square displacement, 
displacement fluctuation, its components parallel and perpendicular to the initial heading direction, and the fourth moment of displacement. They show multiple dynamical crossovers analyzed using exact expressions.  
The calculation of kurtosis identifies deviations from Gaussian behavior at intermediate times. The dynamics is analyzed further by direct numerical simulations calculating the displacement distribution. 
With time, it transforms from an initial unimodal distribution peaked at the origin to a distribution characterizing expanded trajectories via bimodal distributions. Eventually, the expanded state becomes Gaussian with the peak shifting to the origin
via another bimodal distribution characterizing a coexistence. 
The two crossovers via the two bimodalities distinguishes ABPs with speed fluctuations from that with constant speed.

The paper is organized as follows. 
In Sec.~\ref{theory}, we present the model and describe the Laplace transform of Fokker-Planck equation to derive the general expression for dynamical moments in arbitrary dimensions. In Sec.~\ref{lower_order_moments}, we obtain the mean displacement, mean-squared displacement and displacement fluctuations. We demonstrate the anisotropy in displacement fluctuations at short times and analyze their crossovers with time. In Sec.~\ref{fourth_moment_kurtosis}, we calculate the fourth moment of displacement and kurtosis. Using the kurtosis, we show the deviations of the dynamics from Gaussian process. In Sec.~\ref{disp_dist} we use direct numerical simulations to determine the evolution of the probability distribution function of displacement.  Finally, in Sec.~\ref{conclusion}, we conclude by presenting a summary and outlook.

\section{Theory}
\label{theory}

\subsection{Model}
The dynamics of this active particle in $d$-dimensions is described by its position $\rv = (r_1,r_2,\ldots,r_d)$ and orientation $\uv=(u_1,u_2,\ldots,u_d)$, which is a unit vector in $d$-dimensions. Let the infinitesimal increments at time $t$ are denoted by $dr_{i}=r_i(t+dt)-r_i(t)$ and $du_i=u_i(t+dt)-u_i(t)$. Within the Ito convention~\cite{Ito1975,Berg1985,Mijatovic2020}, the equation of motion of the ABP with Gaussian speed fluctuation is given by~\cite{Pietzonka2018},
\begin{align}
dr_i &= (v_0~dt + dB^s )~ u_i +  dB_i^t(t),
\label{eom:disp}
\end{align} 
\begin{align}
du_i &= (\d_{ij}-u_i u_j)\,dB_j^r(t) - (d-1) D_r u_i \,dt,
\label{eom:rot_Ito_Gaussian_speed_dist}
\end{align}
where the translational noise $\bm{dB}^t$ due to the heat bath follows a Gaussian distribution with its components obeying $\la \bm{dB}_{i}^{t} \ra=0$ and $\la dB_i^t dB_j^t  \ra = 2 D \d_{ij} dt$.  
Within a discrete lattice model in Ref.~\cite{Pietzonka2018}, the active displacement was considered to be associated with the release of a chemical potential. In the continuum limit, it led to a speed with deterministic part $v_0$ and speed fluctuations denoted by an additional Gaussian noise $dB^s$ obeying $\la dB^s\ra = 0$ and  $\la dB^s dB^s\ra = 2 \dac\, dt$.  It is easy to see that dimensionally $D_v = \d v^2 \t_v$ with a speed fluctuation $\d v^2$ and an associated relaxation time $\t_v$. Such a relation can be derived directly considering the mechanism of active speed generation~\cite{Schienbein1993, Shee2021b, Pietzonka2018}.  The orientational diffusion of the heading direction is governed by  the Gaussian noise $\bm{dB}^r$ with  its components obeying $\la dB_i^r\ra = 0$ and $\la dB_i^r dB_j^r\ra = 2D_r\d_{ij}\, dt$.  The first term in Eq.(\ref{eom:rot_Ito_Gaussian_speed_dist}) denotes a projection operator for the noise $\bm{dB}^r$ in the $(d-1)$-dimensional plane perpendicular to $d\uv$. The second term ensures the  normalization of the unit vector $\uv^2 = 1 = (\uv+d \uv)^2$. 

It is straightforward to perform direct numerical simulations of Eq.s~(\ref{eom:disp}) and ~(\ref{eom:rot_Ito_Gaussian_speed_dist}) using the Euler-Maruyama integration. The units of time and length are set by $\t_r=1/D_r$ and $\bar\ell = \sqrt{D/D_r}$, respectively. This sets the unit of velocity $\bar{v}=\bar \ell/\t_r = \sqrt{D D_r}.$ 

\subsection{Fokker-Planck equation and calculation of moments} 
The probability distribution $P(\rv,\uv,t)$ of the position $\rv$ and the active orientation $\uv$ of the particle  follows the Fokker-Planck equation
\bea
\p_t P(\rv, \uv, t) &=& \dac (\uv\cdot\nabla)^2 P + D_{r}\nabla_u^2 P + D \nabla^{2} P \nn\\
&-& v_0\, \uv\cdot \nabla P, 
\label{eq:F-P}
\eea
where 
$\nabla$ is the $d$-dimensional Laplacian operator, and $\nabla_u$ is the Laplacian in the ($d-1$) dimensional orientation space and can be expressed as $\nabla^2_u = x^2 \sum_i \p_{x_i}^2 - [x^2 \p_x^2 + (d-1)x \p_x]$ using $u_i = x_i/x$ with $x=|\bm{x}|$. Using the 
Laplace transform $\tilde P(\rv, \uv, s) = \int_0^\infty dt e^{-s t} P(\rv, \uv, t)$ the Fokker-Planck equation becomes,
\bea
-P(\rv, \uv, 0) + s \tilde P(\rv, \uv, s) &=& \dac (\uv\cdot\nabla)^2 \tilde P + D_r \nabla_u^2 \tilde P\nn\\ &+& D \nabla^2 \tilde P - v_0\, \uv\cdot \nabla \tilde P.  \nn
\eea
The mean of the observable $\psi$ in Laplace space $\la \psi \ra_s = \int d\rv \, d\uv\, \psi(\rv, \uv ) \tilde P(\rv, \uv, s)$. Multiplying the above equation by $\psi(\rv, \uv)$ and integrating over all possible $(\rv, \uv)$  we find,
\bea
-\la \psi \ra_0 + s \la \psi \ra_s &=& \dac \la(\uv\cdot\nabla)^2 \psi \ra_s + D_r \la \nabla_u^2 \psi \ra_s\nn\\
&+& D\la\nabla^2 \psi \ra_s + v_0 \, \la \uv\cdot \nabla \psi \ra_s,
\label{moment}
\eea
where, the initial condition sets $\la \psi \ra_0 = \int d\rv \, d\uv\, \psi(\rv, \uv) P(\rv, \uv, 0)$. Without any loss of generality, we consider the initial condition to follow $P(\rv, \uv, 0) = \d(\rv)\d(\uv - \uv_0)$. Eq.~(\ref{moment}) can be utilized to compute all the moments of any dynamical variable in arbitrary dimensions as a function of time. In the following, we consider  moments of displacement and displacement fluctuations characterizing the dynamics.

\section{Displacement}
\label{lower_order_moments}
In Eq.~(\ref{moment}) using $\psi=\uv$ we get $\la \uv \ra_s = \uv_0/(s+(d-1)D_r)$. The mean displacement can be calculated using $\psi = \rv$ in Eq.~(\ref{moment}), along with the expression for $\la \uv \ra_s$ to get  
$\la \rv \ra_s = v_0  \uv_0/s (s+(d-1)D_r)$.
Performing an inverse Laplace transform this leads to
\begin{align}
\la \rv \ra (t) = \f{v_0 \, \uv_0}{(d-1)D_r} \left( 1 - e^{-(d-1)D_r\, t} \right). 
\label{eq_rav}
\end{align}
The mean displacement is independent of the speed fluctuation, as $dB^s$ and $\uv$ are independent stochastic processes and $\la dB^s\ra=0$. This result, thus, is the same as the displacement of ABPs in the absence of speed fluctuations~\cite{Shee2020}.

\subsection{Mean-squared displacement}
\begin{figure}[t]
\begin{center}
\includegraphics[width=8cm]{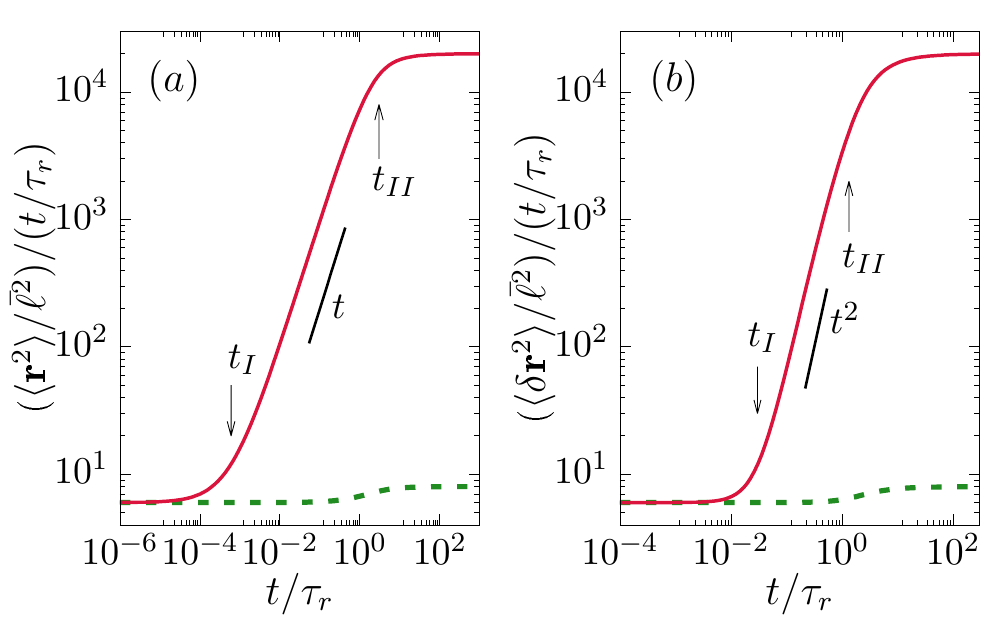}  
\caption{ (color online) Time dependence of $(a)$ $\langle \rv^{2}\rangle$ in Eq.~(\ref{eq_r2d}) and $(b)$ $\langle \delta\rv^{2}\rangle$ in Eq.~(\ref{eq:dr2d})  in $d=2$ for $Pe = v_0/\bar{v} =1$~(dashed line), $100$~(solid line) with $\dact=\dac\t_r/\bar \ell^2=1$. The crossover times for $Pe = 100$ are $(a)$ $t_{I}/\t_r\approx 6\times 10^{-4}$ and $t_{II}/\t_r\approx 3$ and $(b)$ $t_{I}/\t_r = 0.03$ and $t_{II}/\t_r\approx 4/3$.} 
\label{fig:msd_disp_fluct}
\end{center}
\end{figure}

The mean-squared displacement (MSD) can be calculated using $\psi = \rv^2$ in Eq.~(\ref{moment}). With initial position at origin, $\la \rv^2 \ra_0 = 0$. It is easy to see that, $\la \nabla_u^2 \rv^2 \ra_s =0$,  $\la\uv\cdot \nabla \rv^2 \ra_s = 2 \la \uv \cdot \rv \ra_s$, $\la (\uv\cdot\nabla)^2 \rv^2 \ra_s = 2 \la 1 \ra_s$ and $\la \nabla^2 \rv^2 \ra_s = 2 d \la 1 \ra_s$. Note that $\la 1 \ra_s = \int d\rv d\uv \tilde P = \int d\rv d\uv \int_0^\infty dt e^{-st} P=1/s$
using the normalization $\int d\rv d\uv P=1$.
Thus equation~\,(\ref{moment}) leads to 
\bea
s \la \rv^2 \ra_s = 2\dac/s \, +2dD/s \, + 2 v_0 \la \uv \cdot \rv \ra_s. \nn
\eea
We evaluate $\la \uv \cdot \rv \ra_s $ using Eq.~(\ref{moment}) again. Utilizing $\nabla_u^2 \uv = -(d-1)\uv$, $\la \uv \cdot \nabla (\uv\cdot\rv) \ra_s = \la \uv^2 \ra_s = 1/s$, we get 
\bea
\la \uv \cdot \rv\ra_s = v_0/[s(s+(d-1)D_r)].\nn
\eea 
Using this relation in the expression of $\la \rv^2\ra_s$ we obtain
\begin{align}
\la \rv^2\ra_s =  \f{2\dac}{s^2} + \f{2dD}{s^2} + \f{2 v_0^2}{s^2(s+(d-1)D_r)} . 
\label{eq_r2d_Laplace}
\end{align} 
The inverse Laplace transform gives the MSD  
\bea
\la \rv^2 \ra &=& 2 d \left(D+ \f{v_0^2}{(d-1)dD_r}+\f{\dac}{d} \right) t\nn\\
 &-& \f{2 v_0^2}{(d-1)^2D_r^2 } \left(1 -e^{-(d-1)D_r t} \right).
\label{eq_r2d} 
\eea
The time dependence of MSD is shown in Fig.~\ref{fig:msd_disp_fluct}($a$). 
In the long time limit of $D_r t\to \infty$, it gives a  diffusive scaling, 
$\la \rv^2 \ra = 2d\, D_{\rm eff}\, t$ with the effective diffusion constant
\bea
D_{\rm eff} &=& D  + \f{v_0^2}{(d-1)dD_r} + \f{\dac}{d}.
\label{eq:Deff}
\eea
Clearly $D_{\rm eff}$ consists of thermal diffusion $D$, the effective diffusion due to the persistence of motion $v_0^2/[(d-1)d D_r]$ and the contribution from speed fluctuations $D_v/d$.  At $\dac=0$, the expression for $\la \rv^2\ra$ agrees with the results for ABPs in the absence of speed fluctuation~\cite{Shee2020}. Speed fluctuation enhances diffusivity, thereby rendering a mechanism for better spreading which might be utilized, e.g., by pathogenic bacteria in the search of host cells~\cite{Otte2021}.    

\subsubsection{Dynamical crossovers}
In the small time limit of $t\to 0$, expanding Eq.~(\ref{eq_r2d}) around $t=0$ we get 
\begin{align}
\la \rv^2 \ra = 2 d \left(D+\f{\dac}{d} \right) t + v_0^2 t^2 - \f{(d-1)}{3} v_0^2 D_r t^3 +{\cal O}(t^4), \nn
\end{align}
Comparing the consecutive terms in the expansion, we can determine the crossover points shown in Fig.~(\ref{fig:msd_disp_fluct})$(a)$. 
It predicts the first diffusive $\la \rv^2 \ra\sim t$ to ballistic  $\la \rv^2 \ra\sim t^2$ crossover at $t_I \approx 2(dD+\dac)/v_0^2$, followed by a ballistic to diffusive crossover at $t_{II} \approx 3/(d-1)D_r$. In  Fig.~(\ref{fig:msd_disp_fluct})$(a)$, these crossover times are identified for parameter values $\dact=\dac\t_r/\bar \ell^2=1$ and $Pe=100$, they are $t_{I}/\t_r = 2(2+\dact)/Pe^2 \approx 6\times 10^{-4}$ and $t_{II}/\t_r \approx 3$. Similar crossovers are present at small $Pe$ as well, but are less pronounced.

\subsection{Displacement fluctuation}
Using Eq.(\ref{eq_rav}) and Eq.(\ref{eq_r2d}) one can directly obtain the displacement fluctuation $\la \d \rv^2\ra = \la \rv^2\ra - \la \rv \ra^2$  to get 
\bea
\la\d \rv^2 \ra &=& 2 d D_{\rm eff} t 
- \f{v_0^2}{(d-1)^2D_r^2 } \left(3 -4e^{-(d-1)D_r t} \right.\nn\\&+&\left. e^{-2(d-1)D_r t}\right).
\label{eq:dr2d} 
\eea
The time dependence of $\la \d \rv^2\ra$ is plotted in Fig.~(\ref{fig:msd_disp_fluct})$(b)$ at two different $Pe$ values. The plot at large $Pe$ clearly shows a crossover from $\la \d \rv^2\ra \sim t$ to $\la \d \rv^2\ra \sim t^3$ at small $t$, followed by a crossover back to diffusive $\sim t$ scaling at large $t$. This can be understood using the expansion 
\bea
\la \d \rv^2 \ra &=& 2 (dD+\dac) t + \f{2}{3} (d-1) v_0^2 D_r t^3\nn\\
 &-&\f{1}{2} (d-1)^2 v_0^2 D_r^2 t^4 +{\cal O}(t^5). \nn
\eea
It predicts a crossover from diffusive $\la \d \rv^2 \ra\sim t$ scaling to $\la \d \rv^2 \ra\sim t^3$ scaling at $t_I \approx [3(dD+\dac)/(d-1)v_0^2 D_r]^{1/2}$, followed by another possible crossover back to the diffusive scaling near $t_{II} \approx 4/3 (d-1)D_r$. 
In  Fig.~(\ref{fig:msd_disp_fluct})$(b)$, the solid line shows the crossovers 
at $\dact=1$ and $Pe=100$. The figure also shows the estimated crossover times $t_{I}/\t_r = [3(2+\dact)/Pe^2]^{1/2}=0.03$ and $t_{II}/\t_r =4/3$.

\subsection{Components of displacement fluctuation}
\label{sec:components_displacement_fluctuation}
\begin{figure}[!t]
\begin{center}
\includegraphics[width=8cm]{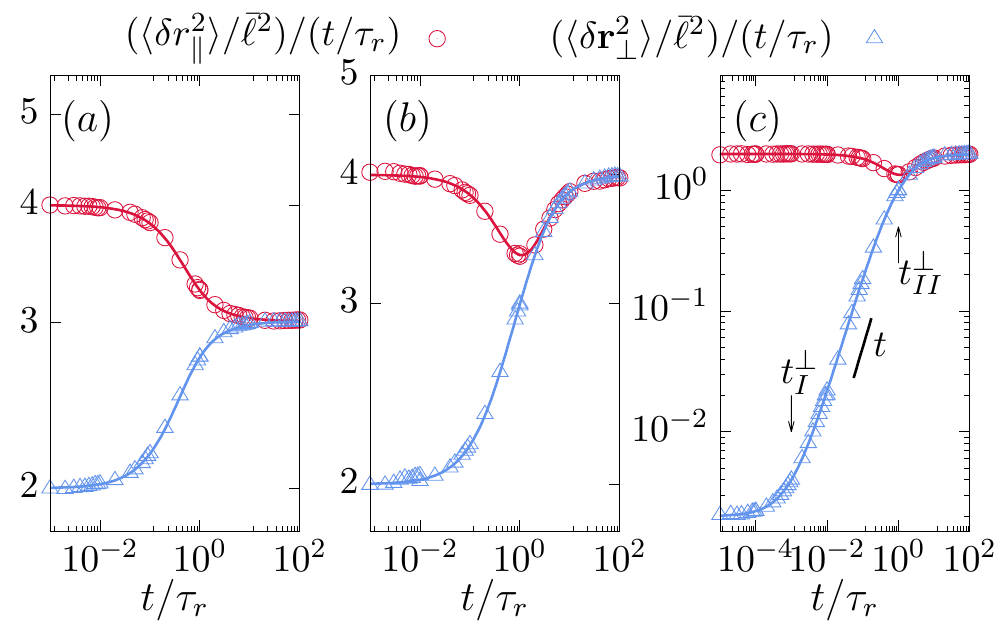} 
\caption{ (color online) Components of displacement fluctuation in two dimensions for low activity $Pe^2\leq \dact$ with $Pe=v_0\t_r/\bar\ell$ and $\dact=\dac\t_r/\bar \ell^2$. Points denote simulations results and lines depict analytical predictions. The components of displacement fluctuations $\langle \delta r_{\parallel}^{2}\rangle$ ($\circ$, red) and $\langle \delta \rv_{\perp}^{2}\rangle$ ($\triangle$, blue) correspond to Eq.~(\ref{eq:drpl2}) and  Eq.~(\ref{eq:drpp2a}) respectively. The parameter values used for $(a)$ $\dact=1$ and $Pe=0.1$, $(b)$ $\dact=1$ and $Pe=1$, $(c)$ $\dact=10^{3}$ and $Pe = 31.62 $. The crossover times in ($c$) are $t_{I}^{\perp}/\t_r=10^{-3}$, and $t_{II}^{\perp}/\t_r=1$. The parallel component shows sub-diffusive behavior at intermediate time-scales as the condition $Pe^2\leq \dact$ is satisfied. } 
\label{fig:comp_disp_fluct1}
\end{center}
\end{figure}

We assume the initial heading direction $\uv_0 = \hat x$ towards the positive $x$-axis. Thus the second moment of the component of displacement parallel to initial heading direction $\rpl^2 = x^2$ can be calculated using $\psi=x^2$ in Eq.(\ref{moment}).  This gives, 
\bea
s \la \rpl^2 \ra_s = 2\dac\la u_x^2 \ra_s \, +2D/s \, + 2v_0 \la x u_x \ra_s. \nn
\eea
Using  Eq.(\ref{moment}) it is straightforward to show $\la u_x^2 \ra_s = \f{(s+2D_r)}{s(s+2 dD_r)}$ and $\la x u_x\ra_s = \f{v_0}{s + (d-1)D_r} \la u_x^2 \ra_s$. 
Thus we obtain
\bea
\la \rpl^2 \ra_s &=& \f{2\dac(s+2D_r)}{s^2(s+2 dD_r)} +\f{2D}{s^2}\nn\\ 
&+& \f{2 v_0^2 (s+ 2D_r)}{s^2(s+ (d-1)D_r )(s+2 dD_r)} .
\eea
Performing the inverse Laplace transform we find the time dependence,
\bea
\la \rpl^2 \ra &=& 2 \left(D +\f{\dac}{d}+ \f{v_0^2}{(d-1)d D_r}\right) t\nn\\
 &+&\f{(d-1)\dac}{d^2 D_r}\left(1-e^{-2 d D_r t}\right)\nn\\
&+& \f{v_0^2}{D_r^2} \left(\frac{(d-1) e^{-2 d D_r t}}{d^2 (d+1)}
+\frac{2 (3-d) e^{-(d-1)D_rt}}{(d-1)^2 (d+1)}\right.\nn\\ &+&\left.\frac{d^2-4 d+1}{(d-1)^2 d^2} \right)
\eea
It is easy to obtain the relative fluctuation $\la \d \rpl^2 \ra = \la \rpl^2 \ra -\la r_{\parallel} \ra^{2}$ noting that the displacement $\la \rpl \ra = \la \rv \cdot \uv_0 \ra = \f{v_0}{(d-1)D_r} \left( 1 - e^{-(d-1)D_r\, t} \right)$.
The fluctuation in the perpendicular component $\la \d \rp^2 \ra = \la \rp^2 \ra$, as  the mean $\la \rp \ra=0$. 
Thus $\la \d \rp^2 \ra = \la \rv^2 \ra - \la \rpl^2\ra$. 
As a result, 
\bea
\la \d \rpl^2 \ra &=& 2 \left(D +\f{\dac}{d}+ \f{v_0^2}{(d-1)d D_r}\right) t\nn\\
 &+&\f{(d-1)\dac}{d^2 D_r}\left(1-e^{-2 d D_r t}\right)\nn\\
&+& \f{v_0^2}{D_r^2} \left(\frac{(d-1) e^{-2 d D_r t}}{d^2 (d+1) } +\frac{8 e^{-(d-1) D_r t}}{(d-1)^2 (d+1) }\right.\nn\\ 
&-&\left.\f{e^{-2(d-1)D_r t}}{(d-1)^2 } -\frac{4 d-1}{(d-1)^2 d^2 } \right), \label{eq:drpl2}
\eea
\bea
\la \d \rp^2 \ra &=& 2 (d-1)\left( D + \f{\dac}{d} + \f{v_0^2}{(d-1)d D_r}\right)\,t\nn\\
&-&\f{(d-1)\dac}{d^2 D_r}\left(1-e^{-2 d D_r t}\right)\nn\\&+& \f{v_0^2}{D_r^2} \left( \f{4 e^{-(d-1)D_r t}}{d^2-1} - \f{(d-1)e^{-2d D_r t}}{d^2(d+1)}\right.\nn\\ &-&\left. \f{3d-1}{d^2(d-1)}\right).
\label{eq:drpp2a}
\eea
In the absence of  speed fluctuation $\dac=0$, the above result reduces to that of  usual ABPs~\cite{Shee2020}.
We show comparisons of direct numerical simulations of the model in $d=2$ with the above-mentioned analytic predictions in Fig.s~\ref{fig:comp_disp_fluct1} and \ref{fig:comp_disp_fluct2}. Remarkably, the parallel component shows a non-monotonic variation in Fig.~\ref{fig:comp_disp_fluct1}. The detailed nature of their time-dependence is further analyzed in the following.

\subsubsection{In two dimensions} 
The above results simplifies in two dimensions, $d=2$. 
In the small time limit expanding the two components around $t=0$ we obtain
\bea
\la \d \rpl^2 \ra_{t\to 0} &=& 2(D+\dac) t -2 \dac D_r t^2 +\f{8}{3} \dac D_r^2 t^3\nn\\
&+& (\f{1}{3} v_0^2 -\f{8}{3} \dac D_r)D_r^2 t^4\nn\\
&-& (\f{7}{15} v_0^2-\f{32}{15}\dac D_r)D_r^3 t^5 + {\cal O}(t^6),
\label{eq:dr2_parallel_small_time}
\eea
\bea
\la \d \rp^2 \ra_{t\to 0} &=& 2Dt + 2 \dac D_r t^2+ (\f{2}{3} v_0^2 -\f{8}{3} \dac D_r)D_r t^3\nn\\
&-& (\f{5}{6} v_0^2 - \f{8}{3}\dac D_r)D_r^2 t^4+ {\cal O}(t^5).
\label{eq:dr2_normal_small_time}
\eea
The resultant small time limit diffusive scalings are $\la \d \rpl^2 \ra_{t\to 0} \approx 2 (D+D_v)t$ and $\la \d \rp^2 \ra_{t\to 0} \approx 2 D t$. Moreover, the above expansions can be used to identify the observed crossovers. Before analyzing them, we note that the components  of displacement fluctuation return to diffusive scaling asymptotically, but with different effective diffusivities 
\begin{align}
&\la \d \rpl^2 \ra_{t\to \infty} = 2 \left(D +\f{\dac}{2}+ \f{v_0^2}{2 D_r}\right) t, \nn\\
&\la \d \rp^2 \ra_{t\to \infty} = 2 \left( D + \f{\dac}{2} + \f{v_0^2}{2 D_r}\right)\,t.
\label{eq:drpp2b}
\end{align} 
The differences between these two limits,
\begin{align}
\la \d \rpl^{2} \ra_{t\to \infty} -\la \d \rpl^2 \ra_{t\to 0} =\left(\f{v_0^2}{D_r}-\dac\right)t,\nn\\
\la \d \rp^{2} \ra_{t\to \infty} -\la \d \rp^2 \ra_{t\to 0} =\left(\f{v_0^2}{D_r}+\dac\right)t, 
\label{eq:comp_diff}
\end{align}
are useful to understand their time dependence. Clearly, $\la \d \rpl^2\ra/t$ will reduce (increase) with time for $v_0^2 < \dac D_r$ ($v_0^2 > \dac D_r$). In contrast, $\la \d \rp^{2} \ra/t$ increases from short time diffusive to asymptotic diffusive behavior, irrespective of the value of active speed.   

\subsubsection{Low activity limit $v_{0}^{2} \leq \dac D_{r}$}
In  Fig.~(\ref{fig:comp_disp_fluct1})$(a)$, the parallel component shows diffusive- subdiffusive- diffusive crossovers. 
In Fig.~(\ref{fig:comp_disp_fluct1})$(b)$ and ($c$), $\la \d \rpl^{2} \ra$ shows diffusive- subdiffusive- super ballistic- diffusive crossovers.  
The crossover points can be estimated by comparing the various $t$-scaling in the right hand side of Eq.(\ref{eq:dr2_parallel_small_time}). The first sub-diffusive crossover appears at $t_I D_r = (D+D_v)/D_v$. The following super-ballistic crossover point to $\la \d \rpl^{2} \ra \sim t^3$ is at $t_{II} D_r=1$. The final diffusive crossover appears at $t_{III}=8D_v/(8 D_v D_r - v_0^2)$. 

In the perpendicular component, 
the crossovers $\la \d \rp^2 \ra\sim t$ to $\la \d \rp^2 \ra\sim t^2$
appears at $t_{I}^{\perp} D_r=D/\dac$. It is followed by a crossover back to $\la \d \rp^2 \ra\sim t$ at $t_{II}^{\perp} D_r\approx [3\dac D_r/(v_0^2-4\dac D_r)]$ if $v_0^2<4\dac D_r$. 
These crossovers are identified in Fig.~(\ref{fig:comp_disp_fluct1})$(c)$ where with $\dact=10^{3}$ and $Pe=31.62$ the condition $Pe^2<4\dact$ holds. The crossover times are $t_{I}^{\perp}\equiv t_{I}^{\perp}/\t_r=1/\dact=10^{-3}$ and $t_{II}^{\perp}\equiv t_{II}^{\perp}/\t_r\approx [3\dact/(Pe^2-4\dact)]=1$.
\begin{figure}[t]
\begin{center}
\includegraphics[width=8cm]{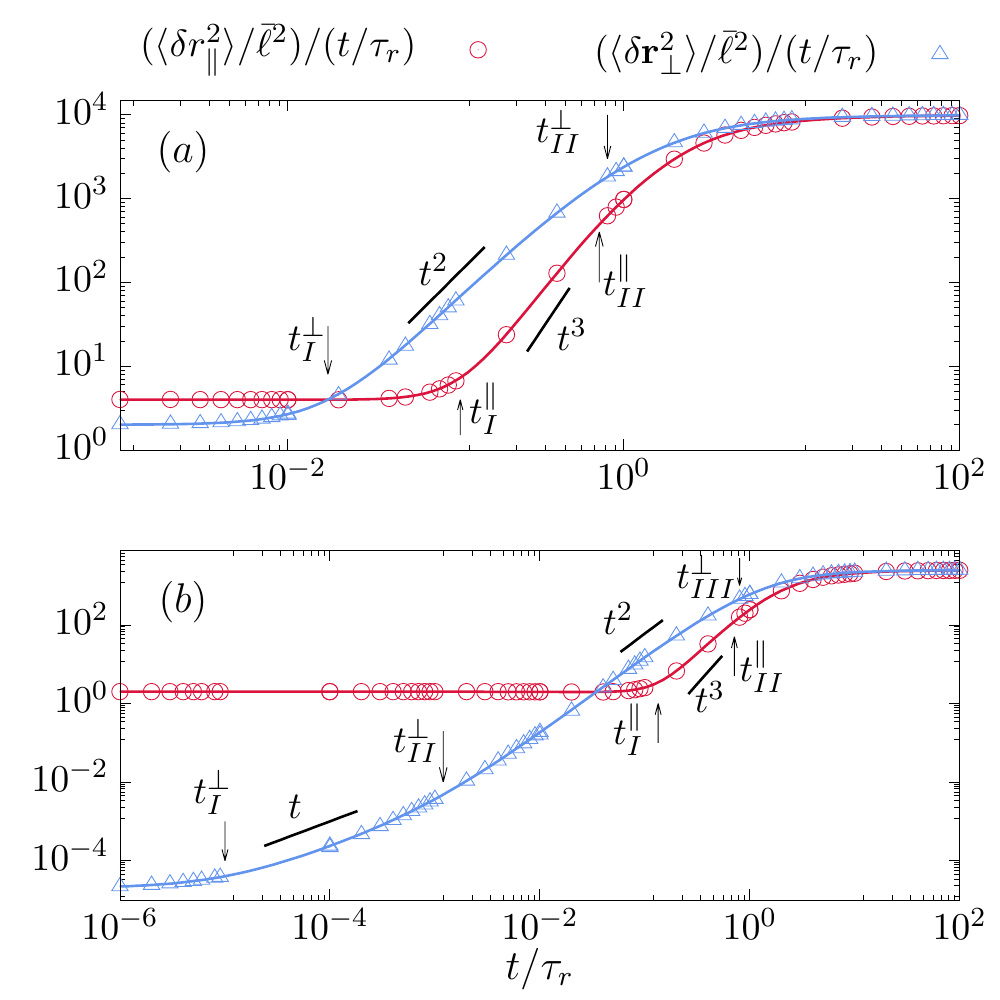} 
\caption{ (color online) Components of displacement fluctuation in $d=2$ for high activity $Pe^2>\dact$ with $Pe=v_0\t_r/\bar\ell$ and $\dact=\dac\t_r/\bar \ell^2$. The points denote numerical simulations and the lines denote analytic expressions. The parallel ($\circ$, red) and perpendicular ($\triangle$, blue)  components of displacement fluctuation correspond to Eq.~(\ref{eq:drpl2}) and  Eq.~(\ref{eq:drpp2a}) respectively. The parameter values used are $(a)$ $\dact=1$, $Pe=10^{2}$ and $(b)$ $\dact=10^{5}$ and $Pe=1.58\times 10^{4}$.  In $(a)$, the crossover times are denoted by $t_{I}^{\parallel}/\t_r=0.11$, $t_{II}^{\parallel}/\t_r=0.71$, $t_{I}^{\perp}/\t_r=1.7\times 10^{-2}$, and $t_{II}^{\perp}/\t_r=0.8$. In $(b)$, the crossover times are denoted by $t_{I}^{\parallel}/\t_r=0.13$, $t_{II}^{\parallel}/\t_r=0.71$, $t_{I}^{\perp}/\t_r= 10^{-5}$, $t_{II}^{\perp}/\t_r=1.2\times 10^{-3}$, and $t_{III}^{\perp}/\t_r=0.8$.} 
\label{fig:comp_disp_fluct2}
\end{center}
\end{figure}
\vskip 0.5cm

\subsubsection{High activity limit $v_{0}^{2} > \dac D_{r}$}
In this limit the final diffusivity in $\la \d \rpl^2 \ra$ can be larger than the short time diffusivity. The parallel component $\la \d \rpl^2 \ra$ first crosses over from $\la \d \rpl^2 \ra\sim t$ to $\la \d \rpl^2 \ra\sim t^3$ at $t_I\approx (3(1+D/\dac)/4)^{1/2}D_r^{-1}$ followed by another crossover from $\la \d \rpl^2 \ra\sim t^3$  to $\la \d \rpl^2 \ra\sim t^4$ at $t_{II}\approx [8\dac D_r/(v_0^2 -8\dac D_r)]D_r^{-1}$ and finally in the long time limit a further crossover to $\la \d \rpl^2 \ra\sim t$ at $t_{III} \approx [5(v_0^2 - 8\dac D_r)/(7v_0^2-32\dac D_r)]D_r^{-1} $ when $t_{I} < t_{II} < t_{III}$ is satisfied. As before, the crossover times are calculated by comparing different terms in Eq.~(\ref{eq:dr2_parallel_small_time}). The condition $t_{III}> t_{II}$ leads to $v_0^2 >((68+\sqrt{1744})/5) \dac D_r$  and the condition $t_{II}>t_{I}$ amounts to $v_0^2<(16(\dac^3/3(D+\dac))^{1/2}+8\dac)D_r$. Even for $D=0$, the condition $t_{II}>t_{I}$ corresponding to  $v_0^2<(16/\sqrt{3}+8)\dac D_r$ conflicts with the assumption of $v_{0}^{2} > \dac D_{r}$. It suggests that $\la \d \rpl^2 \ra\sim t^3$ is not possible.  Thus, the possible crossovers are $\la \d \rpl^2 \ra\sim t$ to $\la \d \rpl^2 \ra\sim t^4$, finally to $\la \d \rpl^2 \ra\sim t$. The first crossover $\la \d \rpl^2 \ra\sim t$ to $\la \d \rpl^2 \ra\sim t^4$ can appear at $t_{I}^{\parallel}=\left[6(D+\dac)/(v_{0}^2 - 8 \dac D_{r})\right]^{1/3}$ and the second crossover $\la \d \rpl^2 \ra\sim t^4$ to $\la \d \rpl^2 \ra\sim t$ can appear at $t_{II}^{\parallel}=t_{III}=[5(v_0^2 - 8\dac D_r)/(7v_0^2-32\dac D_r)]D_r^{-1}$. 

In Fig.~(\ref{fig:comp_disp_fluct2}), we show the crossovers $\la \d \rpl^2 \ra\sim t$ to $\sim t^4$ finally to $\sim t$. The crossover times in Fig.~(\ref{fig:comp_disp_fluct2})$(a)$ for $\dact=1$ and $Pe=10^2$ are $t_{I}^{\parallel}/\t_r=[6(1+\dact)/(Pe^2 - 8 )]^{1/3}\approx 0.11$ and $t_{II}^{\parallel}/\t_r=[5(Pe^2 - 8\dact )/(7Pe^2-32\dact )]\approx 0.71$. Similarly, the crossover times in Fig.~(\ref{fig:comp_disp_fluct2})$(b)$ for $\dact=10^5$ and $Pe=1.58\times 10^4$ are $t_{I}^{\parallel}/\t_r \approx 0.13$ and $t_{II}^{\parallel}/\t_r\approx 0.71$.

One possible scenario of crossovers in $\la \d \rp^2 \ra$ is the following: (i)~from $\la \d \rp^2 \ra\sim t$ to $\la \d \rp^2 \ra\sim t^3$ at $t_{I}^{\perp\prime} D_r=[3DD_r/(v_0^2 - 4 \dac D_r)]^{1/2} $ with the condition $v_0^2>4\dac D_r+3\dac^2 D_r/D$ ($t_{II}^{\perp}<t_{I}^{\perp}$), (ii)~back to $\la \d \rp^2 \ra\sim t$ at $t_{III}^{\perp} D_r = [4(v_0^2 -4 \dac D_r)/(5v_0^2-16\dac D_r)]$ if the condition $v_0^2>[(47+\sqrt{417})/8]\dac D_r$($t_{II}^{\perp}<t_{III}^{\perp}$) is satisfied. Moreover,  $t_{III}>t_{I}$ leads to the condition $v_0^2 > 16(D-\dac)\dac D_r/(5D-4\dac)$. The crossovers $\la \d \rp^2 \ra\sim t$ to $\sim t^3$ and finally to $\sim t$ are shown in Fig.~(\ref{fig:comp_disp_fluct2})$(a)$ for $\dact = 1$, $Pe= 10^2$. The crossover times identified in the figure are $t_{I}^{\perp} \equiv t_{I}^{\perp}/\t_r=[3/(Pe^2 - 4 \dact)]^{1/2} =1.7\times 10^{-2}$, $t_{II}^{\perp}\equiv t_{II}^{\perp}/\t_r = [4(Pe^2 -4 \dact)/(5 Pe^2-16\dact )]=0.8$

Another scenario of possible crossovers are $\la \d \rp^2 \ra\sim t$ to $\la \d \rp^2 \ra\sim t^2$ at $t_{I}^{\perp} D_r=D/\dac$ with condition $v_0^2<4\dac D_r+3\dac^2 D_r/D$ ($t_{II}^{\perp}>t_{I}^{\perp}$) to $\la \d \rp^2 \ra\sim t^3$ at $t_{II}^{\perp} = 3\dac/(v_0^2-4\dac D_r)$ with condition $v_0^2>[(47+\sqrt{417})/8]\dac D_r$($t_{III}^{\perp}>t_{II}^{\perp}$) to $\la \d \rp^2 \ra\sim t$ with condition $v_0^2>4\dac D_r$ at $t_{III}^{\perp} D_r = [4(v_0^2 -4 \dac D_r)/(5v_0^2-16\dac D_r)]$. These are shown in Fig.~(\ref{fig:comp_disp_fluct2})$(b)$ at  $\dact=10^5$ and $Pe=1.58\times 10^4$. The identified crossover times are $t_{I}^{\perp}\equiv t_{I}^{\perp}/\t_r=1/\dact=10^{-5}$, $t_{II}^{\perp}\equiv t_{II}^{\perp}/\t_r = 3\dact/(Pe^2-4\dact)\approx 1.2\times 10^{-3}$ and $t_{III}^{\perp}\equiv t_{III}^{\perp}/\t_r = 4(Pe^2 -4 \dact)/(5 Pe^2-16\dact)\approx 0.8$.

\section{Fourth moment and kurtosis}
\label{fourth_moment_kurtosis}
In this section we obtain the fourth moment of displacement $\la \rv^4 \ra$ and hence the kurtosis to quantify  the deviations from possible Gaussian behavior.  
Proceeding as before, using $\psi = \rv^4$ in Eq.~(\ref{moment}) and the relations
\bea
 s \la \rv^4 \ra_s &=& 4 \dac (\la \rv^2 \ra_s  + 2\la (\uv\cdot\rv)^2 \ra_s)+ 4(d+2)D \la\rv^2\ra_s\nn\\&+& 4 v_0 \la (\uv \cdot \rv) \rv^2 \ra_s, \nn\\
 s \la\uv\cdot\rv\ra_s &=& -(d-1)D_r \la\uv\cdot\rv\ra_s + v_0 \la 1\ra_s,\nn\\
 s \la (\uv \cdot \rv)^2 \ra_s &=& 2\dac\la 1 \ra_s + 2 D_r \la r^2\ra_s - 2dD_r \la (\uv \cdot \rv)^2 \ra_s\nn\\ &+&2D\la 1 \ra_s+ 2 v_0 \la \uv \cdot \rv \ra_s,\nn\\
s \la (\uv \cdot \rv) \rv^2  \ra_s &=& 6 \dac \la \uv \cdot \rv\ra_s - (d-1) D_r \la (\uv \cdot \rv) \rv^2  \ra_s\nn\\ &+&(4+2d)D\la \uv \cdot \rv\ra_s + v_0 \la \rv^2 \ra_s + 2 v_0 \la (\uv \cdot \rv)^2 \ra_s\nn
\eea
it is straightforward to obtain  the fourth moment of displacement in the Laplace space  
\bea
\la \rv^4 \ra_s &=& 8[\dac + (d+2)D](\dac +dD) \frac{1 }{s^3} + \f{16\dac(\dac+D)}{s^2 (s+2dD_r)}\nn\\&+& \f{32\dac (\dac +D)D_r}{s^3 (s+2dD_r)}+\f{8\dac v_0^2 (5s+2(d-1)D_r)}{s^3 (s+(d-1)D_r)^2}\nn\\
&+&\f{32\dac v_0^2 (s+2D_r)}{s^3 (s+(d-1)D_r)(s+2dD_r)}\nn\\&+&\f{8D v_0^2(d+2)(3s + 2 (d-1) D_r)}{s^3 (s+(d-1)D_r)^2}\nn\\ &+&  \f{8 v_0^4 (3s + 2 (d+2) D_r)}{s^3 (s+(d-1) D_r)^2 (s + 2d D_r)}.
\label{eq:r4s}
\eea
\begin{widetext} 
Performing the inverse Laplace transform, we obtain the time evolution of the fourth moment, 
\begin{align}
&\la \rv^4(t) \ra = 4 [\dac + (d+2)D](\dac +dD) t^2 + 16 \dac (\dac + D)\left[\f{t}{2dD_r}-\f{1}{(2dD_r)^2}\left(1-e^{-2dD_r t}\right)\right]\nn\\
&+32\dac (\dac + d D) D_r\left[\f{t^2}{4dD_r}-\f{t}{(2dD_r)^2}+\f{1}{(2dD_r)^3}\left(1-e^{-2dD_r t}\right)\right]\nn\\
&+8\dac v_0^2\left[\f{t^2}{(d-1)D_r} +\f{t}{(d-1)^2 D_r^2}+\f{3te^{-(d-1)D_r t}}{(d-1)^2 D_r^2}-\f{4}{(d-1)^3 D_r^3}\left(1-e^{-(d-1)D_r t}\right)\right]\nn\\
&+32\dac v_0^2 \left[\f{t^2}{2(d-1)dD_r}+\f{(d^2 -4d+1)t}{2(d-1)^2 d^2 D_r^2}+\f{-3d^3+11d^2-5d+1}{4(d-1)^3 d^3 D_r^3}+\f{(d-3)e^{-(d-1)D_r t}}{(d-1)^3 (d+1)D_r^3}-\f{(d-1)e^{-2dD_r t}}{4d^3 (d+1)D_r^3}\right]\nn\\
&-\frac{8 \left(d^2 v_0^4+10 d v_0^4+25 v_0^4\right) e^{-(d-1) D_r t}}{(d-1)^4 (d+1)^2 D_r^4}
+\frac{4 \left(d^3 v_0^4+23 d^2 v_0^4-7 d v_0^4+v_0^4\right)}{(d-1)^4 d^3 D_r^4}\nn\\
&+\frac{8 t e^{- (d-1) D_r t } \left(d^3 D D_r v_0^2+2 d^2 D D_r v_0^2-d D D_r v_0^2+d v_0^4-2 D D_r v_0^2-7 v_0^4\right)}{(d-1)^3 (d+1) D_r^3}\nn\\
&   +\frac{4 t^2 \left(d^5 D^2 D_r^2-3 d^3 D^2 D_r^2+2 d^3 D D_r v_0^2+2 d^2 D^2 D_r^2+2 d^2 D D_r v_0^2-4 d D D_r v_0^2+d v_0^4+2 v_0^4\right)}{(d-1)^2 d D_r^2}\nn\\
 &  -\frac{8 t \left(d^4 D D_r v_0^2+d^3 D D_r v_0^2-2 d^2 D D_r v_0^2+d^2 v_0^4+6 d v_0^4-v_0^4\right)}{(d-1)^3 d^2 D_r^3} .
\label{r4t}
\end{align}
\end{widetext}

\begin{figure}[!t]
\begin{center}
\includegraphics[width=8cm]{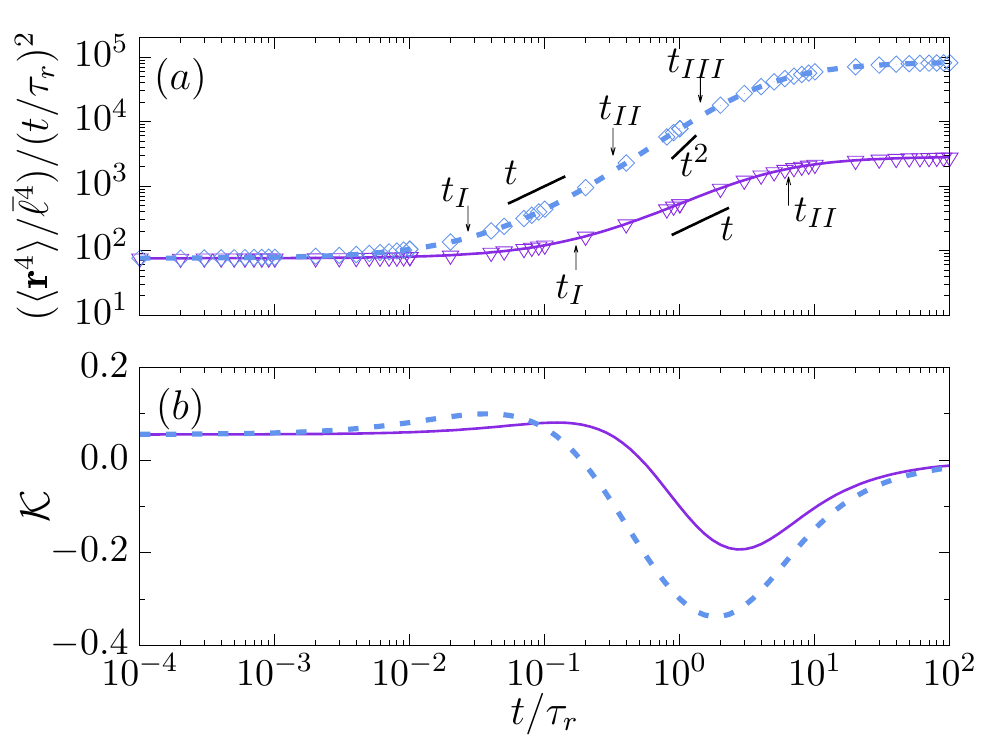}
\caption{ (color online) $(a)$ Fourth moment and $(b)$ kurtosis of displacement as a function of time in $d=2$ at $\dact=1$. ($a$)~The points denote simulation results and the lines are plots of Eq.(\ref{r4t}). At $Pe=v_{0}\t_r/\bar\ell=4~(\triangledown)$ two crossovers are identified at $t_{I}/\t_r\approx0.17$ and $t_{II}/\t_r\approx 6.38$. The plot for  $Pe=10~(\Diamond)$ shows three crossovers at $t_{I}/\t_r\approx 0.027$, $t_{II}/\t_r\approx 0.32$, and $t_{III}/\t_r\approx 1.42$. $(b)$~Plot of kurtosis ${\cal K}$ as a function of time at $Pe=4$ (solid line) and $Pe=10$ (dashed line).} 
\label{fig:r4avg_kurtosis2}
\end{center}
\end{figure}

For $\dac=0$ this result agrees with the fourth moment of usual ABPs as obtained in Ref.~\cite{Shee2020}. 
The fourth moment of a general Gaussian process obeys 
\begin{align}
\mu_4 =\la \rv^2 \ra^2 + \f{2}{d}\left(\la \rv^2 \ra^2 -  \la \rv \ra^4\right).
\end{align}
Using the expression of $\la \rv^4 \ra(t)$, the kurtosis in $d$-dimensions is defined as 
\bea
{\cal K} = \f{\la \rv^4 \ra}{\mu_4} - 1 .
\eea
In Fig.\ref{fig:r4avg_kurtosis2}($a$) we show the comparison between analytic expression (lines) and numerical simulation results  (points) in $d=2$ dimensions for $\la \rv^4\ra$. Fig.\ref{fig:r4avg_kurtosis2}($b$) shows the time-dependence of kurtosis. To analyze the crossovers in $\la \rv^4\ra$ in $d=2$, we expand the analytical expression in Eq.(\ref{r4t}) around $t=0$  to obtain
\begin{align}
 &\la \rv^4(t) \ra = (12\dac^2 +32\dac D + 32 D^2) t^2 \nn\\
 &+ \left[4(3\dac+4D) v_0^2  -\f{16}{3} \dac^2 D_r\right] t^3\nn\\
 & + \left(v_0^4 +\f{16\dac^2 D_r^2}{3} -   \f{16}{3} D D_r v_0^2-   \f{20}{3} \dac D_r v_0^2  \right) t^4 \nn\\
 &- \left(\f{2}{3} v_0^4 D_r  +\f{64\dac^2 D_r^3}{15}- \f{4 D D_r^2 v_0^2}{3} -\f{11\dac D_r^2 v_0^2}{3}\right) t^5 \nn\\
 &+ {\cal O}(t^6).
 \label{r4td2_expansion}
\end{align}

\begin{figure}[!t]
\begin{center}
\includegraphics[width=8cm]{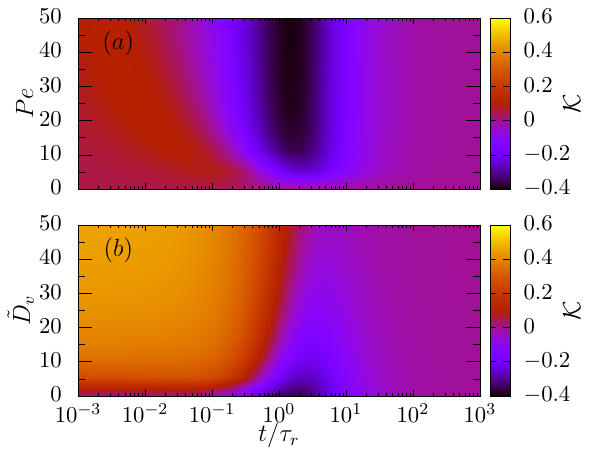}
\caption{ (color online) Deviation from Gaussian nature: kymographs of the kurtosis $\mathcal{K}$ as a function of time $t/\t_r$ of the two-dimensional ABP for different $Pe$ at $\dact = 1$~($a$) and for different $\dact$ at $Pe=10$~($b$).} 
\label{fig:kurtosis_map}
\end{center}
\end{figure}

This gives the expression in short time limit 
\bea
\la \rv^4\ra_{t\to 0} = (12\dac^2 +32\dac D + 32 D^2) t^2.
\label{eq_r40} 
\eea
In the long time limit, Eq.(\ref{r4t}) for $d=2$ gives 
\bea
\la \rv^4(t) \ra_{t\to \infty} \approx 8\left[\left(\dac+2D\right)^2  +\f{2\dac v_{0}^2}{D_r}\right] t^2
\label{eq_r4inf}
\eea
The difference between the small and long time fourth order moments gives
 \bea
\la \rv^{4} \ra_{t\to \infty} -\la \rv^4 \ra_{t\to 0} \approx \f{4\dac}{D_r}\left(4 v_0^2 -\dac D_r\right)t^2,
\eea
which arises due to the active speed fluctuation. Whether $\la \rv^4\ra$ will eventually increase or decrease with time depends on if $4 v_0^2>D_v D_r$  ($Pe^2>\dact/4$) or $4 v_0^2 < D_v D_r$  ($Pe^2 < \dact/4$). The parameter values in Fig.\ref{fig:r4avg_kurtosis2}($a$) obey the condition $Pe^2>\dact/4$ and thus shows increase in $\la \rv^4\ra$ with time. 

The expansion in Eq.(\ref{r4td2_expansion}) shows that at short time the scaling $ \la \rv^4(t) \ra \sim t^2$ can cross over to $ \la \rv^4(t) \ra \sim t^3$  at $t_I = 3(3\dac^2 +8D(\dac+D))/[3(3\dac + 4D)v_0^2 -4\dac^2 D_r]$ provided $v_0^2 > 4\dac^2 D_r/ [3 (3 \dac + 4 D)]$ or $Pe^2 > \f{4}{3}\f{\dact^2}{3 \dact +4}$. The nature of the crossovers that follow depends on the activity and can be analyzed using the expansion in Eq.(\ref{r4td2_expansion}). 
The solid line in Fig.~(\ref{fig:r4avg_kurtosis2})$(a)$ corresponding to $\dact=1$ and $Pe=4$ shows $\la \rv^4(t) \ra \sim t^2$ to $\sim t^3$ crossover at $t_{I}/\t_r=3(3\dact^2 +8(1+\dact))/[3(3\dact + 4) Pe^2 -4\dact^2]\approx 0.17$, followed by a crossover back to $\sim t^2$ at $t_{II}/\t_r = 4[3(3\dact+4)Pe^2 -4\dact^2]/[3Pe^4 + 16\dact^2 -4(4 + 5\dact) Pe^2]\approx 6.38$. 
The dashed line in Fig.~(\ref{fig:r4avg_kurtosis2})$(a)$  corresponding to $\dact=1$ and $Pe=10$ shows the first crossover from $\la \rv^4(t) \ra \sim t^2$ to $\sim t^3$  at $t_{I}/\t_r=3(3\dact^2 +8(1+\dact))/[3(3\dact + 4) Pe^2 -4\dact^2]\approx 0.027$. The second crossover from $\la \rv^4(t) \ra \sim t^3$ to $\sim t^4$ appears at $t_{II}/\t_r = 4[3(3\dact+4)Pe^2 -4\dact^2]/[3Pe^4 + 16\dact^2 -4(4 + 5\dact) Pe^2]\approx 0.32$. The final crossover $\la \rv^4(t) \ra \sim t^4$ to $\sim t^2$ appears at $t_{III} \equiv t_{III}/\t_r = 5[3 Pe^4 +16 \dact^2 -4 (4 + 5\dact) Pe^2]/[10 Pe^4 +64 \dact^2 -5(4 + 11\dact) Pe^2]\approx 1.42$.

To characterize the deviation from a possible Gaussian behavior, e.g., as expected in the active Ornstein-Uhlenbeck process~\cite{Fodor2016, Das2018}, we show the evolution of kurtosis as a function of time in Fig.~\ref{fig:r4avg_kurtosis2}$(b)$. The deviation from the Gaussian behavior shows up in the form of a positive kurtosis in the short time regime governed by the speed fluctuation. 
Eventually the kurtosis shows an intermediate time deviations to negative values, controlled by the orientational fluctuations,  before asymptotically vanishing corresponding to a long time Gaussian limit. 

In Fig.~(\ref{fig:kurtosis_map})($a$), we show a kymograph of kurtosis describing its time evolution at different $Pe$, keeping the speed fluctuation $\dact=1$ fixed. The amount of negative deviation of ${\cal K}$ at intermediate times increases with $Pe$ to eventually saturate at ${\cal K} \approx -0.4$.  At larger $Pe$, the deviations towards negative kurtosis appear earlier in time. At longer times, ${\cal K}$ vanishes asymptotically.  Fig.~(\ref{fig:kurtosis_map})($b$) shows the kymograph of ${\cal K}$ describing its time evolution at different $\dact$ for a fixed $Pe=10$. At short times ${\cal K}$ remains positive, showing increased positive deviations in the presence of larger speed fluctuation $\dact$. Again, ${\cal K}$ shows deviations to negative values at intermediate times before vanishing asymptotically. However, the onset of negative deviations of kurtosis requires longer time in the presence of stronger speed fluctuations.

\section{Displacement distribution}

\begin{figure}[!t]
\begin{center}
\includegraphics[width=8cm]{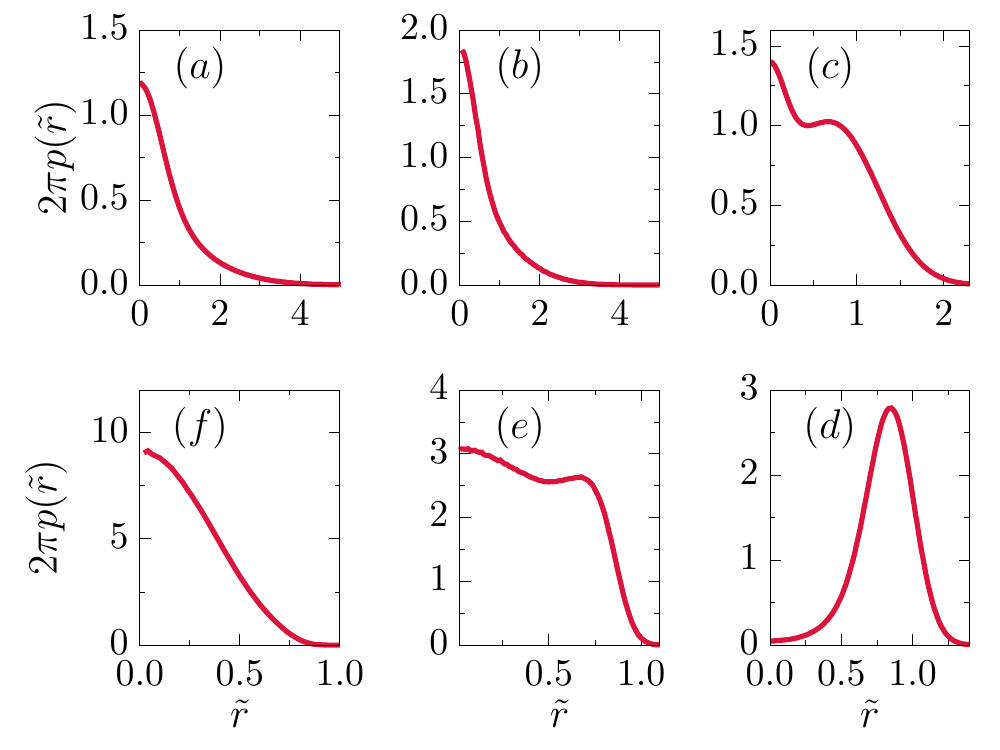} 
\caption{ (color online) Probability distributions of displacement $2\pi p(\tilde{r})$ with $\tilde{r}=r/L$ at $Pe = v_{0}\t_r/\bar \ell = 31.6$ and $\tilde{\dac}=\dac\t_r/\bar \ell^2=10$ over different time-segments expressed as $\tilde L= v_0 t/\bar \ell = 0.32~(a)$, $0.63~(b)$, $3.16~(c)$, $31.62~(d)$, $126.49~(e)$, and $316.23~(f)$. 
}
\label{fig_Dist}
\end{center}
\end{figure}

To gain further insights into the dynamical crossovers, we present displacement distributions obtained from direct numerical simulations for ABP trajectories of dimensionless length $\tilde L=L/\bar \ell$ where $L=v_0 t$. In Fig.~\ref{fig_Dist} we plot the distribution functions $p(\tilde{r})$ of the scaled displacement $\tilde{r}=r/L$ at $Pe = v_{0}\t_r/\bar \ell = 31.6$ and $\tilde{\dac}=\dac\t_r/\bar \ell^2=10$. With increasing length of trajectories the distribution transforms from a unimodal distribution with maximum at $\tilde r=0$ in Fig.~\ref{fig_Dist}($a$) to one with the maximum corresponding to  extended trajectories with $\tilde r \approx 1$ in Fig.~\ref{fig_Dist}($d$) to finally a Gaussian distribution with the maximum at $\tilde r=0$  in Fig.~\ref{fig_Dist}($f$). Both the transformations between extended and compact trajectories are mediated by bimodal distributions as can be seen in Fig.s~\ref{fig_Dist}($c$) and ($e$).   

Note that the control parameters $\dact$ and $Pe$ can be expressed in terms of $D$, $D_p=v_0^2/D_r$ and $\dac$, the three terms controlling the effective diffusion in Eq.(\ref{eq:Deff}), such that $\dact=\dac/D$ and $Pe^2=D_p/D$. Thus the relative strength of  $Pe$ and $\dact^{1/2}$ influences the displacement statistics. Further, the  ratio $\tilde L/Pe$ is equivalent to the persistence ratio $L/\l$ of the trajectory length $L=v_0 t$ and the persistence length $\l=v_0 \t_r$. This ratio is known to control the extension statistics of persistent random walks and worm like chains~\cite{Shee2020, Dhar2002}. As can be seen from Fig.~\ref{fig_Dist}, the value of dimensionless trajectory length $\tilde L$ compared to the speed fluctuation scale $\dact^{1/2}$ and the activity $Pe$ determines the properties of the displacement distributions.

\label{disp_dist}

In Fig.~\ref{fig_Dist}($a$) and Fig.~\ref{fig_Dist}($b$), $\tilde L \ll \dact^{1/2} < Pe$. From the perspective of directional persistence, the trajectories in this regime are equivalent to rigid rods, as the persistence ratio $\tilde L/Pe = L/\l \sim 10^{-2}$. The unimodal distribution $p(\tilde r)$ with the maximum at $\tilde r=0$ for these quasi-one dimensional trajectories are determined by the speed fluctuation $\dact$ alone. The increased fluctuation due to $\dact$ in Fig.~\ref{fig_Dist}($b$) shrinks the trajectories further producing a narrower distribution $p(\tilde r)$.  This behavior changes into a bimodal distribution in Fig.~\ref{fig_Dist}($c$) where $\tilde L\sim \dact^{1/2} < Pe$. The maximum at the origin is again due to the speed fluctuations. However, with respect to the trajectory length the speed fluctuation is significantly smaller than the previous two cases, allowing the system to show the second maximum in $p(\tilde r)$ near $\tilde r \approx 1$ corresponding to extended trajectories of persistent motion at $\tilde L/Pe=0.1$. For longer trajectories in Fig.~\ref{fig_Dist}($d$)--($f$), $\tilde L > \dact^{1/2}$, the speed fluctuation can be neglected and the change in $p(\tilde r)$ can be interpreted in terms of simple persistent motion and equivalently the WLC polymer~\cite{Dhar2002,Shee2020}. The single maximum in $p(\tilde r)$ in  Fig.~\ref{fig_Dist}($d$) corresponds to extended configurations of a WLC polymer at persistent ratio $\tilde L/Pe=1$. Similar behavior was observed earlier in Ref.~\cite{Dhar2002,Shee2020}. Fig.~\ref{fig_Dist}($e$) corresponds to the persistent ratio $\tilde L/Pe=4$. The bistability observed in this regime is equivalent to the rigid rod- flexible chain bistability observed for WLCs in the same regime of persistent ratio~\cite{Dhar2002}. For longer trajectories with $\tilde L/Pe=10$, the distribution turns into unimodal Gaussian distribution with the maximum at $\tilde r=0$. This is the asymptotic long time behavior of the trajectories and correspond to the flexible chain limit of WLCs~\cite{Dhar2002}. 

Note that the first crossover from contracted trajectories in  Fig.~\ref{fig_Dist}($a$) to extended trajectories in Fig.~\ref{fig_Dist}($d$) via the bimodality in Fig.~\ref{fig_Dist}($c$) is due to the active speed fluctuation. This behavior is absent in ABPs moving with constant speed. The second crossover from the extended state in Fig.~\ref{fig_Dist}($d$) to the Gaussian contracted state  in Fig.~\ref{fig_Dist}($f$) is controlled by persistence as the impact of speed fluctuations for these long trajectories can be neglected. In a recent publication we have shown a mapping of trajectories of ABPs with constant speed and in the presence of thermal diffusion to configurations of a semiflexible polymer~\cite{Shee2020}.  Thus the second crossover seen in the present context is similar to the transition in polymer properties in the WLC model via phase coexistence.

\section{Discussion} 
\label{conclusion} 
We considered the impact of active speed fluctuations on a $d$-dimensional active Brownian particle (ABP). We utilized a Laplace transform method for the Fokker-Planck equation, originally proposed to understand the worm-like chain (WLC) model of semiflexible polymers~\cite{Hermans1952}, to find exact expressions for dynamical moments of ABPs in arbitrary dimensions. This method allowed us to obtain several such moments, including the mean-squared displacement, displacement fluctuations parallel and perpendicular to the initial heading direction, and the fourth moment of displacement to characterize the dynamics.
We found several dynamical crossovers and identified the crossover points 
using the exact analytic expressions.
They depend on the activity, persistence, and speed fluctuation of the ABP.

The persistence in the motion led to an anisotropy captured by the parallel and perpendicular components of the displacement fluctuation with respect to the initial heading direction. As we   showed, the parallel component can display sub-diffusive behavior and non-monotonic variations at intermediate times, unlike the perpendicular component. The exact calculation of kurtosis measuring the non-Gaussian nature of the stochastic displacement showed positive values at short times controlled by the speed fluctuation. 
It crossed over to a negative minimum at intermediate times, a behavior governed by the persistence of motion, before vanishing asymptotically at long times characterizing the asymptotic Gaussian nature of the ABP trajectories. 

To further analyze the dynamics, we used direct numerical simulations in two dimensions to obtain the probability distributions of ABP displacement as the time elapsed.  It showed two crossovers between contracted and expanded trajectories via two separate bimodal distributions at intervening times. 
The first crossover from contracted to expanded trajectories showed a clear bimodality at intermediate times signifying coexistence 
of the two kinds of trajectories.This crossover is determined by the speed fluctuation and is absent in ABPs with constant speed.
The second crossover between expanded and contracted trajectories appearing at later times was controlled by the persistence. 
Such a crossover mediated via bimodal distributions is equivalent to the transition between the rigid rod and flexible polymer via the coexistence of the two conformational phases observed in the WLC model~\cite{Dhar2002}. 

The generation of active speed from underlying stochastic mechanisms, e.g., as considered in Ref.~\cite{Pietzonka2018, Schienbein1993, Schweitzer1998, Romanczuk2012}, involves inherent speed fluctuations. Such fluctuations are present in active colloids performing phoretic motion~\cite{Bechinger2016}  and mechanisms generating motion in motile cells and bacteria~\cite{Marchetti2013, Wu2009, Theves2013}. Our predictions can be tested in experiments on tagged active particles, and our results can be used in analyzing the dynamics of motile cells.  In a dense dispersion of ABPs, inter-particle collisions can effectively enhance active speed fluctuations~\cite{Redner2013,Fily2014}. In their run and tumble motion, several bacteria show switching between active speeds~\cite{PerezIpina2019, Otte2021}. Our methods can be extended to understand the non-equilibrium dynamics of such systems better.

\acknowledgements

The numerical calculations were supported in part by SAMKHYA, the
high performance computing facility at Institute of Physics, Bhubaneswar. We thank Abhishek Dhar for discussions on a related project. D.C. thanks SERB, India for financial support through grant number MTR/2019/000750 and International Centre for Theoretical Sciences (ICTS) for an associateship.

\bibliographystyle{prsty.bst}

\begin{thebibliography}{10}

\bibitem{Vicsek2012}
T. Vicsek and A. Zafeiris, Phys. Rep. {\bf 517},  71  (2012).

\bibitem{Romanczuk2012}
P. Romanczuk, M. B{\"{a}}r, W. Ebeling, B. Lindner, and L. Schimansky-Geier,
  European Physical Journal: Special Topics {\bf 202},  1  (2012).

\bibitem{Astumian2002}
R.~D. Astumian and P. H{\"{a}}nggi, Physics Today {\bf 55},  33  (2002).

\bibitem{Reimann2002}
P. Reimann, Physics Report {\bf 361},  57  (2002).

\bibitem{Berg1972}
H.~C. Berg and D.~A. Brown, Nature {\bf 239},  500  (1972).

\bibitem{Niwa1994}
H.~S. Niwa, Journal of Theoretical Biology {\bf 171},  123  (1994).

\bibitem{Ginelli2015}
F. Ginelli, F. Peruani, M.-H. Pillot, H. Chat{\'{e}}, G. Theraulaz, and R. Bon,
  Proc. Natl. Acad. Sci. {\bf 112},  12729  (2015).

\bibitem{Marchetti2013}
M.~C. Marchetti, J.~F. Joanny, S. Ramaswamy, T.~B. Liverpool, J. Prost, M. Rao,
  and R.~A. Simha, Reviews of Modern Physics {\bf 85},  1143  (2013).

\bibitem{Bechinger2016}
C. Bechinger, R. {Di Leonardo}, H. L{\"{o}}wen, C. Reichhardt, G. Volpe, and G.
  Volpe, Reviews of Modern Physics {\bf 88},  045006  (2016).

\bibitem{Bar2020}
M. B{\"{a}}r, R. Gro{\ss}mann, S. Heidenreich, and F. Peruani, Annu. Rev.
  Condens. Matter Phys. {\bf 11},  441  (2020).

\bibitem{Das2018}
S. Das, G. Gompper, and R.~G. Winkler, New Journal of Physics {\bf 20},  015001
   (2018).

\bibitem{Kurzthaler2018b}
C. Kurzthaler, C. Devailly, J. Arlt, T. Franosch, W.~C. Poon, V.~A. Martinez,
  and A.~T. Brown, Physical Review Letters {\bf 121},  078001  (2018).

\bibitem{Malakar2018}
K. Malakar, V. Jemseena, A. Kundu, K. {Vijay Kumar}, S. Sabhapandit, S.~N.
  Majumdar, S. Redner, and A. Dhar, Journal of Statistical Mechanics: Theory
  and Experiment {\bf 2018},  43215  (2018).

\bibitem{Basu2018a}
U. Basu, S.~N. Majumdar, A. Rosso, and G. Schehr, Physical Review E {\bf 98},
  062121  (2018).

\bibitem{Peruani2007}
F. Peruani and L.~G. Morelli, Physical Review Letters {\bf 99},  010602
  (2007).

\bibitem{Basu2019}
U. Basu, S.~N. Majumdar, A. Rosso, and G. Schehr, Physical Review E {\bf 100},
  062116  (2019).

\bibitem{Majumdar2020}
S.~N. Majumdar and B. Meerson, Phys. Rev. E {\bf 102},  022113  (2020).

\bibitem{Wagner2017}
C.~G. Wagner, M.~F. Hagan, and A. Baskaran, Journal of Statistical Mechanics:
  Theory and Experiment {\bf 2017},  043203  (2017).

\bibitem{Elgeti2015}
J. Elgeti, R.~G. Winkler, and G. Gompper, Reports on progress in physics.
  Physical Society (Great Britain) {\bf 78},  056601  (2015).

\bibitem{Dhar2019}
A. Dhar, A. Kundu, S.~N. Majumdar, S. Sabhapandit, and G. Schehr, Phys. Rev. E
  {\bf 99},  032132  (2019).

\bibitem{Shee2020}
A. Shee, A. Dhar, and D. Chaudhuri, Soft Matter {\bf 16},  4776  (2020).

\bibitem{Santra2020a}
I. Santra, U. Basu, and S. Sabhapandit, Phys. Rev. E {\bf 101},  062120
  (2020).

\bibitem{Pototsky2012}
A. Pototsky and H. Stark, Europhysics Letters {\bf 98},  50004  (2012).

\bibitem{Malakar2020}
K. Malakar, A. Das, A. Kundu, K.~V. Kumar, and A. Dhar, Physical Review E {\bf
  101},  22610  (2020).

\bibitem{Chaudhuri2021}
D. Chaudhuri and A. Dhar, Journal of Statistical Mechanics: Theory and
  Experiment {\bf 2021},  013207  (2021).

\bibitem{Howse2007}
J.~R. Howse, R.~A. Jones, A.~J. Ryan, T. Gough, R. Vafabakhsh, and R.
  Golestanian, Physical Review Letters {\bf 99},  048102  (2007).

\bibitem{Palacci2010}
J. Palacci, C. Cottin-Bizonne, C. Ybert, and L. Bocquet, Physical Review
  Letters {\bf 105},  088304  (2010).

\bibitem{Redner2013}
G.~S. Redner, M.~F. Hagan, and A. Baskaran, Physical Review Letters {\bf 110},
  055701  (2013).

\bibitem{Fily2014}
Y. Fily, A. Baskaran, and M.~F. Hagan, Soft Matter {\bf 10},  5609  (2014).

\bibitem{Sevilla2014}
F.~J. Sevilla and L.~A. {G{\'{o}}mez Nava}, Physical Review E {\bf 90},  022130
   (2014).

\bibitem{Basu2018}
U. Basu, S.~N. Majumdar, A. Rosso, and G. Schehr, Physical Review E {\bf 98},
  1  (2018).

\bibitem{Wu2009}
Y. Wu, A.~D. Kaiser, Y. Jiang, and M.~S. Alber, Proceedings of the National
  Academy of Sciences {\bf 106},  1222  (2009).

\bibitem{Theves2013}
M. Theves, J. Taktikos, V. Zaburdaev, H. Stark, and C. Beta, Biophys. J. {\bf
  105},  1915  (2013).

\bibitem{PerezIpina2019}
E. {Perez Ipi{\~{n}}a}, S. Otte, R. Pontier-Bres, D. Czerucka, and F. Peruani,
  Nat. Phys. {\bf 15},  610  (2019).

\bibitem{Otte2021}
S. Otte, E.~P. Ipi{\~{n}}a, R. Pontier-Bres, D. Czerucka, and F. Peruani,
  Nature Communications {\bf 12},  1990  (2021).

\bibitem{Schweitzer1998}
F. Schweitzer, W. Ebeling, and B. Tilch, Phys. Rev. Lett. {\bf 80},  5044
  (1998).

\bibitem{Dadhichi2018}
L. {Prawar Dadhichi}, A. Maitra, and S. Ramaswamy, J. Stat. Mech. Theory Exp.
  {\bf 2018},  123201  (2018).

\bibitem{Ramaswamy2017}
S. Ramaswamy, J. Stat. Mech. Theory Exp. {\bf 2017},  054002  (2017).

\bibitem{Pietzonka2018}
P. Pietzonka and U. Seifert, Journal of Physics A: Mathematical and Theoretical
  {\bf 51},  01LT01  (2018).

\bibitem{Pietzonka2019}
P. Pietzonka, {\'{E}}. Fodor, C. Lohrmann, M.~E. Cates, and U. Seifert, Phys.
  Rev. X {\bf 9},  41032  (2019).

\bibitem{Pietzonka2016}
P. Pietzonka, K. Kleinbeck, and U. Seifert, New Journal of Physics {\bf 18},
  052001  (2016).

\bibitem{Hermans1952}
J.~J. Hermans and R. Ullman, Physica {\bf 18},  951  (1952).

\bibitem{Ito1975}
K. It{\^o},  in {\em International Symposium on Mathematical Problems in
  Theoretical Physics}, edited by H. Araki (Springer-Verlag,
  Berlin-Heidelberg-New York, 1975), Chap.~Stochastic Calculus, pp.\ 218--223.

\bibitem{Berg1985}
M. van~den Berg and J.~T. Lewis, Bull. London Math. Soc. {\bf 17},  144
  (1985).

\bibitem{Mijatovic2020}
A. Mijatovi{\'c}, V. Mramor, and G.~U. Bravo, Statistics \& Probability Letters
  {\bf 165},  108836  (2020).

\bibitem{Schienbein1993}
M. Schienbein and H. Gruler, Bulletin of Mathematical Biology {\bf 55},  585
  (1993).

\bibitem{Shee2021b}
A. Shee and D. Chaudhuri, arXiv:2108.12228  (2021).

\bibitem{Fodor2016}
{\'{E}}. Fodor, C. Nardini, M.~E. Cates, J. Tailleur, P. Visco, and F. {Van
  Wijland}, Physical Review Letters {\bf 117},  38103  (2016).

\bibitem{Dhar2002}
A. Dhar and D. Chaudhuri, Physical Review Letters {\bf 89},  065502  (2002).

\end{thebibliography}

\end{document}